\documentclass[10pt,letter,prl,twocolumn,showpacs]{revtex4}
\usepackage{ifthen}
\newboolean{uprightparticles}
\setboolean{uprightparticles}{false} 

\usepackage{amssymb}
\usepackage{graphicx}
\usepackage{amsfonts}
\usepackage{upgreek}
\usepackage{xspace}
\usepackage{amsmath}

\def\lhcb {LHCb\xspace}
\def\ux85 {UX85\xspace}

\def\lhc {LHC\xspace}

\def\rich   {RICH\xspace}

\ifthenelse{\boolean{uprightparticles}}%
{

 \def\Pmu         {\ensuremath{\upmu}\xspace}

 \def\PDelta      {\ensuremath{\Delta}\xspace}                 
 \def\PXi      {\ensuremath{\Xi}\xspace}                 
 \def\PLambda      {\ensuremath{\Lambda}\xspace}                 
 \def\PSigma      {\ensuremath{\Sigma}\xspace}                 
 \def\POmega      {\ensuremath{\Omega}\xspace}                 
 \def\PUpsilon      {\ensuremath{\Upsilon}\xspace}                 
 
 \def\PB      {\ensuremath{\mathrm{B}}\xspace}                 
                  
 \def\PD      {\ensuremath{\mathrm{D}}\xspace}

 \def\PK      {\ensuremath{\mathrm{K}}\xspace}

 \def\PW      {\ensuremath{\mathrm{W}}\xspace}

 \def\Pc      {\ensuremath{\mathrm{c}}\xspace}

 \def\Pi      {\ensuremath{\mathrm{i}}\xspace}

}
{

 \def\Pmu         {\ensuremath{\mu}\xspace}

 \mathchardef\PDelta="7101
 \mathchardef\PXi="7104
 \mathchardef\PLambda="7103
 \mathchardef\PSigma="7106
 \mathchardef\POmega="710A
 \mathchardef\PUpsilon="7107
                  
 \def\PB      {\ensuremath{B}\xspace}                 
                  
 \def\PD      {\ensuremath{D}\xspace}

 \def\PK      {\ensuremath{K}\xspace}

 \def\PW      {\ensuremath{W}\xspace}

 \def\Pc      {\ensuremath{c}\xspace}

 \def\Pi      {\ensuremath{i}\xspace}

}

\def\mup        {\ensuremath{\Pmu^+}\xspace}
\def\mun        {\ensuremath{\Pmu^-}\xspace} 

\def\W      {\ensuremath{\PW}\xspace}

\def\c     {\ensuremath{\Pc}\xspace}

\def\kaon  {\ensuremath{\PK}\xspace}
  \def\Kbar  {\kern 0.2em\overline{\kern -0.2em \PK}{}\xspace}

\def\Kz    {\ensuremath{\kaon^0}\xspace}
\def\Kzb   {\ensuremath{\Kbar^0}\xspace}
\def\KzKzb {\ensuremath{\Kz \kern -0.16em \Kzb}\xspace}
\def\Kp    {\ensuremath{\kaon^+}\xspace}
\def\Km    {\ensuremath{\kaon^-}\xspace}

\def\KpKm  {\ensuremath{\Kp \kern -0.16em \Km}\xspace}

  \def\Dbar    {\kern 0.2em\overline{\kern -0.2em \PD}{}\xspace}
\def\D       {\ensuremath{\PD}\xspace}

\def\Dz      {\ensuremath{\D^0}\xspace}
\def\Dzb     {\ensuremath{\Dbar^0}\xspace}
\def\DzDzb   {\ensuremath{\Dz {\kern -0.16em \Dzb}}\xspace}
\def\Dp      {\ensuremath{\D^+}\xspace}
\def\Dm      {\ensuremath{\D^-}\xspace}

\def\DpDm    {\ensuremath{\Dp {\kern -0.16em \Dm}}\xspace}

\def\Dstarm  {\ensuremath{\D^{*-}}\xspace}

\def\Ds      {\ensuremath{\D^+_s}\xspace}

\def\Dssm     {\ensuremath{\D^{*-}_s}\xspace}

\def\Dsm     {\ensuremath{\D^-_s}\xspace}

\def\B       {\ensuremath{\PB}\xspace}
  \def\Bbar    {\kern 0.18em\overline{\kern -0.18em \PB}{}\xspace}

\def\Bd      {\ensuremath{\B^0}\xspace}
\def\Bs      {\ensuremath{\B^0_s}\xspace}

  \def\Y#1S{\ensuremath{\PUpsilon{(#1S)}}\xspace}

\def\BR         {{\ensuremath{\cal B}\xspace}}

\newcommand{\decay}[2]{\ensuremath{#1\!\to #2}\xspace}         

\def\to                 {\ensuremath{\rightarrow}\xspace}

\def\AT#1     {\ensuremath{A_T^{#1}}\xspace}           

\def\Bsmm     {\decay{\Bs}{\mup\mun}\xspace}

\def\C#1      {\ensuremath{\mathcal{C}_{#1}}\xspace}                       
\def\Cp#1     {\ensuremath{\mathcal{C}_{#1}^{'}}\xspace}                    
\def\Ceff#1   {\ensuremath{\mathcal{C}_{#1}^{\mathrm{(eff)}}}\xspace}        
\def\Cpeff#1  {\ensuremath{\mathcal{C}_{#1}^{'\mathrm{(eff)}}}\xspace}       
\def\Ope#1    {\ensuremath{\mathcal{O}_{#1}}\xspace}                       
\def\Opep#1   {\ensuremath{\mathcal{O}_{#1}^{'}}\xspace}                    

\newcommand{\tev}{\ensuremath{\mathrm{\,Te\kern -0.1em V}}\xspace}
\newcommand{\gev}{\ensuremath{\mathrm{\,Ge\kern -0.1em V}}\xspace}
\newcommand{\mev}{\ensuremath{\mathrm{\,Me\kern -0.1em V}}\xspace}
\newcommand{\kev}{\ensuremath{\mathrm{\,ke\kern -0.1em V}}\xspace}
\newcommand{\ev}{\ensuremath{\mathrm{\,e\kern -0.1em V}}\xspace}
\newcommand{\gevc}{\ensuremath{{\mathrm{\,Ge\kern -0.1em V\!/}c}}\xspace}
\newcommand{\mevc}{\ensuremath{{\mathrm{\,Me\kern -0.1em V\!/}c}}\xspace}
\newcommand{\gevcc}{\ensuremath{{\mathrm{\,Ge\kern -0.1em V\!/}c^2}}\xspace}
\newcommand{\gevgevcccc}{\ensuremath{{\mathrm{\,Ge\kern -0.1em V^2\!/}c^4}}\xspace}
\newcommand{\mevcc}{\ensuremath{{\mathrm{\,Me\kern -0.1em V\!/}c^2}}\xspace}

\def\invpb {\ensuremath{\mbox{\,pb}^{-1}}\xspace}

\newcommand{\chisq}{\ensuremath{\chi^2}\xspace}

\def\gsim{{~\raise.15em\hbox{$>$}\kern-.85em
          \lower.35em\hbox{$\sim$}~}\xspace}
\def\lsim{{~\raise.15em\hbox{$<$}\kern-.85em
          \lower.35em\hbox{$\sim$}~}\xspace}

\def\pt         {\mbox{$p_T$}\xspace}

\def\tell1  {TELL1\xspace}
\def\ukl1   {UKL1\xspace}

\newcommand{\ie}{\mbox{\itshape i.e.}}

\newcommand{\BdDp}{\ensuremath{\Bd      \to \Dm       \pi^+\,}}
\newcommand{\BdDK}{\ensuremath{\Bd      \to \Dm         K^+\,}}
\newcommand{\BsDp}{\ensuremath{\Bs      \to \Dsm     \pi^+\,}}

\newcommand{\BdDstarp}{\ensuremath{\Bd  \to \Dstarm    \pi^+\,}}
\newcommand{\BdDstarK}{\ensuremath{\Bd  \to \Dstarm      K^+\,}}
\newcommand{\BdDrho}{  \ensuremath{\Bd  \to \Dm      \rho^+\,}}
\newcommand{\LbLcp}{\ensuremath{\Lambda_b      \to \Lambda_c^+ \pi^-\,}}

\newcommand{\BsDstarp}{\ensuremath{\Bs  \to \Dssm    \pi^+\,}}

\newcommand{\BsDrho}{  \ensuremath{\Bs  \to \Dsm      \rho^+\,}}

\newcommand{\Dsp}{\ensuremath{\Dsm \pi^+}}
\newcommand{\DdK}{\ensuremath{\Dm K^+}}
\newcommand{\Ddp}{\ensuremath{\Dm \pi^+}}

\newcommand{\fsfdt}{\ensuremath{f_s/f_d}~}

\usepackage{hyperref}
\usepackage[all]{hypcap}

\begin{document}
\title{Determination of $\boldsymbol{f_s/f_d}$ for 7~TeV $\boldsymbol{pp}$ collisions
and a measurement of the branching fraction of the decay $\boldsymbol{\Bd \to \Dm K^+}$}
\author{R.~Aaij et al., (The LHCb Collaboration)}
\begin{abstract}
  \noindent
The relative abundance of the three decay modes \BdDK, \BdDp and \BsDp
produced in 7~TeV $pp$ collisions at the LHC is determined from data
corresponding to an integrated luminosity of $35 \invpb$.
The branching fraction of \BdDK is found to be 
$\BR\left(\BdDK\right) = (2.01 \pm 0.18^{\textrm{stat}} \pm 0.14^{\textrm{syst}})\times 10^{-4}$.
The ratio of fragmentation fractions \fsfdt is determined through the relative abundance of \BsDp to
\BdDK and \BdDp, leading to  
$\fsfdt = 0.253 \pm 0.017 \pm 0.017 \pm 0.020$, where the uncertainties are
statistical, systematic, and theoretical respectively.
\end{abstract}
\maketitle
\noindent Knowledge of the production rate of \Bs mesons is required to determine any \Bs branching fraction.
This rate is determined by the $b\bar{b}$ production cross-section and the fragmentation probability $f_s$, which is the 
fraction of \Bs mesons amongst all weakly-decaying bottom hadrons.
Similarly the production rate of \Bd mesons is driven by the fragmentation probability $f_d$. 
The measurement of the branching fraction of the rare decay \Bsmm is a prime example where improved knowledge of
\fsfdt is needed to reach the highest sensitivity in the search for 
physics beyond the Standard Model~\cite{Aaij:2011Bsmm_notitle}. 
The ratio \fsfdt is in principle dependent on collision energy and
type as well as the acceptance region of the detector. This is the
first measurement of this quantity at the LHC.

The ratio \fsfdt can be extracted if the ratio of branching fractions of 
$B^0$ and $B_s^0$ mesons decaying to particular
final states $X_1$ and $X_2$, respectively, is known:
\begin{equation}
\frac{f_s}{f_d} = 
\frac{N_{X_{2}}}{N_{X_{1}}} 
\frac{     \BR(B^0 \rightarrow X_1)}{     \BR(B_s^0 \rightarrow X_2)}  
\frac{\epsilon(B^0 \rightarrow X_1)}{\epsilon(B_s^0 \rightarrow X_2)}.
\end{equation}
The ratio of the branching fraction of the \BsDp and \BdDK decays is dominated by contributions
from colour-allowed tree-diagram amplitudes
and is therefore theoretically well understood. In contrast, the ratio
of the branching ratios of the two decays \BsDp and \BdDp can be measured with a smaller statistical uncertainty
due to the greater yield of the \Bd mode, but suffers
from an additional theoretical uncertainty due to the contribution
from a \W-exchange diagram.
Both ratios are exploited here to measure \fsfdt 
according to the equations~\cite{Fleischer:2010-2_notitle,Fleischer:2010ay_notitle}
\begin{equation}\label{eq:fs-det1}
\frac{f_s}{f_d}= 0.971 \cdot \left|\frac{V_{us}}{V_{ud}} \right|^2 \left( \frac{f_K}{f_{\pi}} \right)^2 \frac{\tau_{B_d}}{\tau_{B_s}} \frac{1}{{\cal N}_a {\cal N}_F}
\frac{\epsilon_{\DdK}}{\epsilon_{\Dsp}}
\frac{N_{\Dsp}} {N_{\DdK}},
\end{equation} 
and
\begin{equation}\label{eq:fs-det2}
\frac{f_s}{f_d}= 0.982  \cdot \frac{\tau_{B_d}}{\tau_{B_s}} \frac{1}{{\cal N}_a {\cal N}_F {\cal N}_E}
\frac{\epsilon_{\Ddp}}{\epsilon_{\Dsp}}
\frac{N_{\Dsp}} {N_{\Ddp}}.
\end{equation}
Here $\epsilon_X$ is the selection efficiency of decay $X$ (including the branching fraction of the \D decay
mode used to reconstruct it), $N_X$ is the observed
number of decays of this type, the $V_{ij}$ are elements of the CKM matrix, $f_i$ are the meson decay constants and the numerical factors take into account the phase space difference for the ratio of the two decay modes. 
Inclusion of charge conjugate modes is implied throughout. 
The term ${\cal N}_a$ parametrizes non-factorizable SU(3)-breaking effects; ${\cal N}_F$ is
the ratio of the form factors; ${\cal N}_E$ is an additional correction term to account
for the \W-exchange diagram in the \BdDp decay.
Their values~\cite{Fleischer:2010ay_notitle,Fleischer:2010-2_notitle} are
${\cal N}_a=1.00 \pm 0.02$, ${\cal N}_F = 1.24 \pm 0.08$, and ${\cal N}_E = 0.966 \pm 0.075$.
The latest world average~\cite{HFAG_notitle} is used
for the \B meson lifetime ratio $\tau_{B_s}/\tau_{B_d} = 0.973 \pm 0.015$. The numerical values used for the other factors are:
$|V_{us}|=0.2252$, $|V_{ud}|=0.97425$, $f_{\pi}=130.41$ and $f_{K}=156.1$, with 
negligible associated uncertainties~\cite{PDG_2010_notitle}. 

The observed yields of these three decay modes in  $35 \invpb$ of data collected with the \lhcb
detector in the 2010 running period are used to measure \fsfdt averaged
over the \lhcb acceptance and to improve
the current measurement of the branching fraction of the \BdDK decay mode~\cite{Abe:2001waa_notitle}.  

The \lhcb experiment~\cite{DetectPaper_notitle} is a single-arm spectrometer,
designed to study \B decays at the \lhc, with a pseudorapidity acceptance of $2<\eta<5$ for charged tracks.
The first trigger level allows the selection of events with \B hadronic decays using
the transverse energy of hadrons measured in the calorimeter system. 
The event information is subsequently sent to a software trigger, implemented
in a dedicated processor farm, which performs a final online selection of events
for later offline analysis. 
The tracking system determines the momenta of \B decay products
with a precision of $\delta p/p =0.35$--$0.5\%$.
Two Ring Imaging Cherenkov (\rich) detectors
allow charged kaons and pions to be distinguished in the momentum range 2--100~\gevc~\cite{DetectPaper}.

The three decay modes,
$B^0\to D^-(K^+\pi^-\pi^-) \pi^+$, $B^0\to D^-(K^+\pi^-\pi^-) K^+$ and $B^0_s\to D_s^-(K^+K^-\pi^-) \pi^+$,
are topologically identical and can therefore be selected using identical
geometric and kinematic criteria, thus minimizing efficiency differences between them. 
Events are selected at the first trigger stage by requiring a hadron with transverse energy greater than
$3.6$~GeV in the calorimeter.
The second, software, stage~\cite{Gligorov:1300771_notitle,hlt2toponote_notitle} requires
a two, three, or four track secondary
vertex with a high sum \pt of the tracks, 
significant displacement from the primary interaction,
and at least one track with exceptionally
high \pt, large displacement from the primary interaction, and small fit \chisq.

The decays of \B mesons
can be distinguished from background using variables such as the \pt and 
impact parameter \chisq of the \B, \D, and the final state particles
with respect to the primary interaction. 
In addition the vertex quality of
the \B and \D candidates, the \B lifetime, and the angle between the \B momentum vector and
the vector joining the \B production and decay vertices are used in the selection.
The \D lifetime and flight distance are not used in the selection 
because the lifetimes of the \Dsm and \Dm differ by about a factor of two.

The event sample is first selected
using the gradient boosted decision tree
technique~\cite{TMVA_notitle}, which combines the geometrical and
kinematic variables listed above.
The selection is trained on a mixture of simulated 
\BdDp~decays and combinatorial
background selected from the sidebands of the data mass distributions.
The distributions of the input variables for data and simulated signal events show
excellent agreement, justifying the use of simulated events in the training
procedure.

Subsequently, \Dm (\Dsm) candidates are identified by requiring the invariant mass under the 
$K\pi\pi$ ($KK\pi$) hypothesis to fall within the selection 
window $1870^{+24}_{-40}$ ($1969^{+24}_{-40}$)~\mevcc, where the
mass resolution is approximately $10$~\mevcc.
The final \BdDp and \BsDp subsamples consist of events that pass 
a particle identification (PID) criterion on the bachelor particle, based on the 
difference in log-likelihood between the charged pion and kaon hypotheses (DLL)
of $\textrm{DLL}(K-\pi)<0$, with an efficiency of $83.0\%$.
The \BdDK subsample consists of events with $\textrm{DLL}(K-\pi)>5$,
with an efficiency of $70.2\%$. 
Events not satisfying either condition are not used. 

The relative efficiency of the selection procedure is evaluated for all
decay modes using simulated events, where the appropriate resonances in the charm decays are taken into account. 
As the analysis is only sensitive to relative
efficiencies, the impact of differences between data and simulation is small. The relative efficiencies 
are $\epsilon_{\Ddp}/\epsilon_{\DdK} = 1.221 \pm 0.021$, $\epsilon_{\DdK}/\epsilon_{\Dsp}  = 0.917 \pm 0.020$,
and $\epsilon_{\Ddp}/\epsilon_{\Dsp} = 1.120 \pm 0.025$,
where the errors are due to the limited size of the simulated event samples.

\begin{figure}[!b]
 \centering
\includegraphics[width=.4\textwidth]{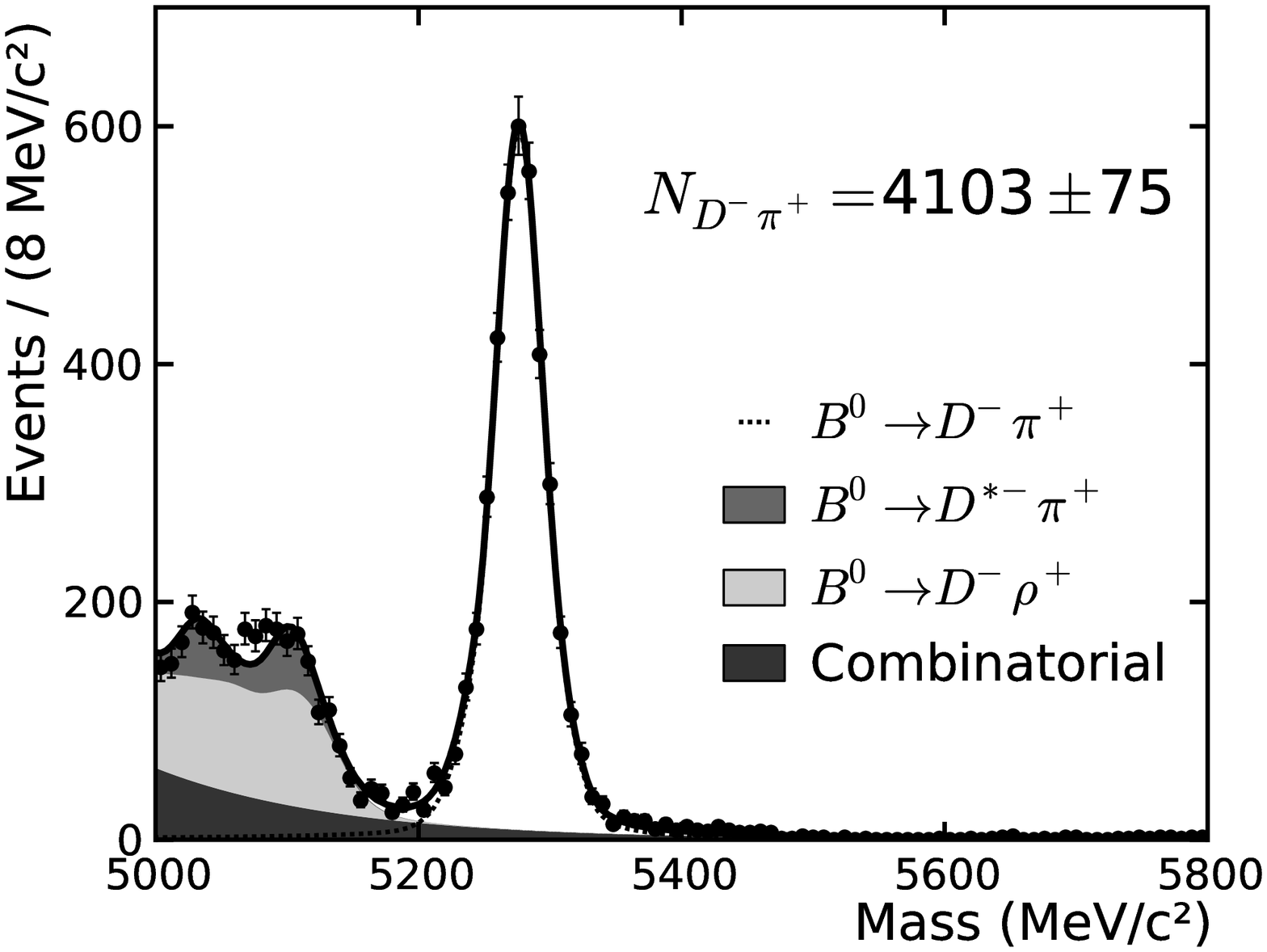}
\includegraphics[width=.4\textwidth]{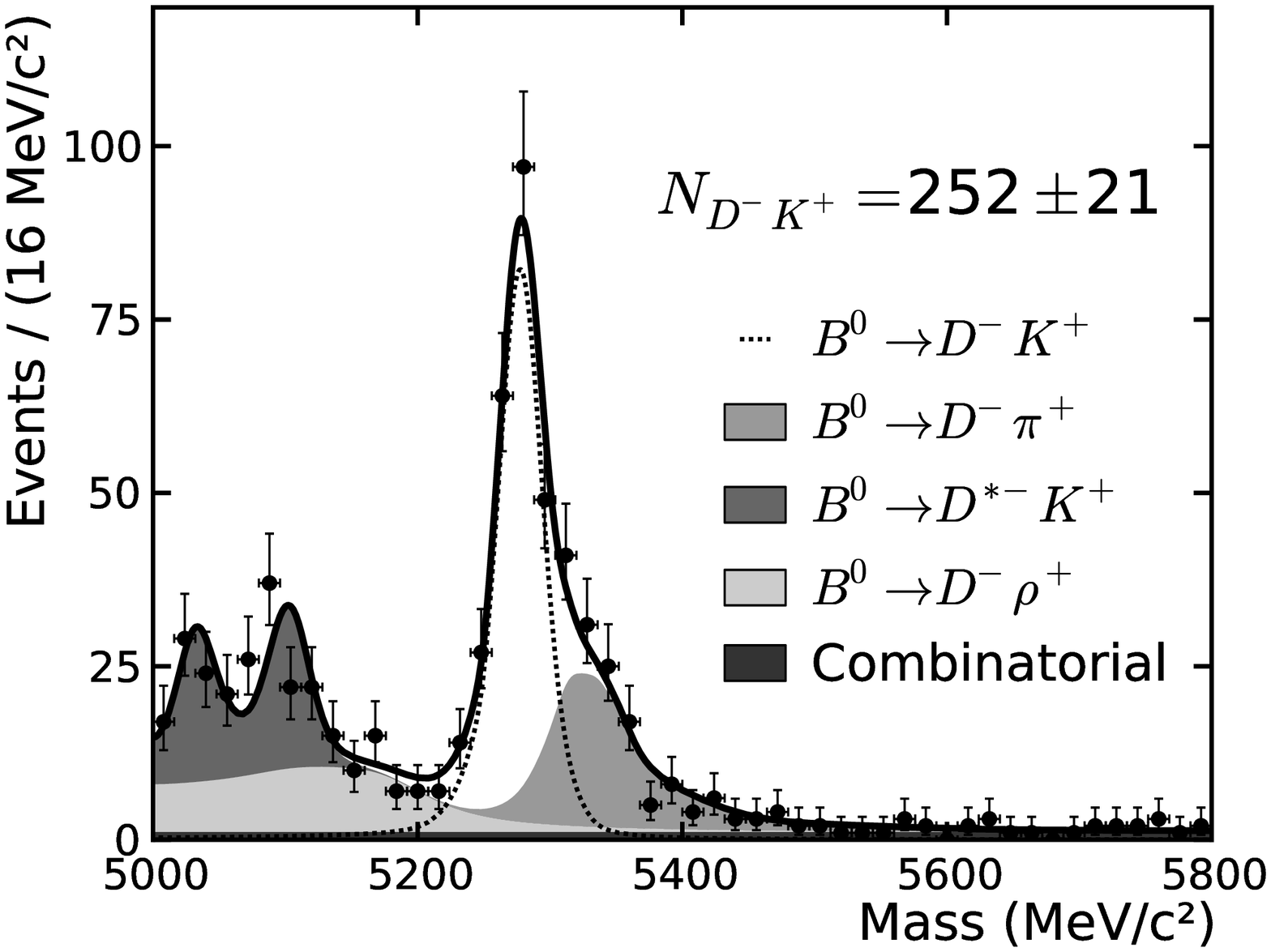}
\includegraphics[width=.4\textwidth]{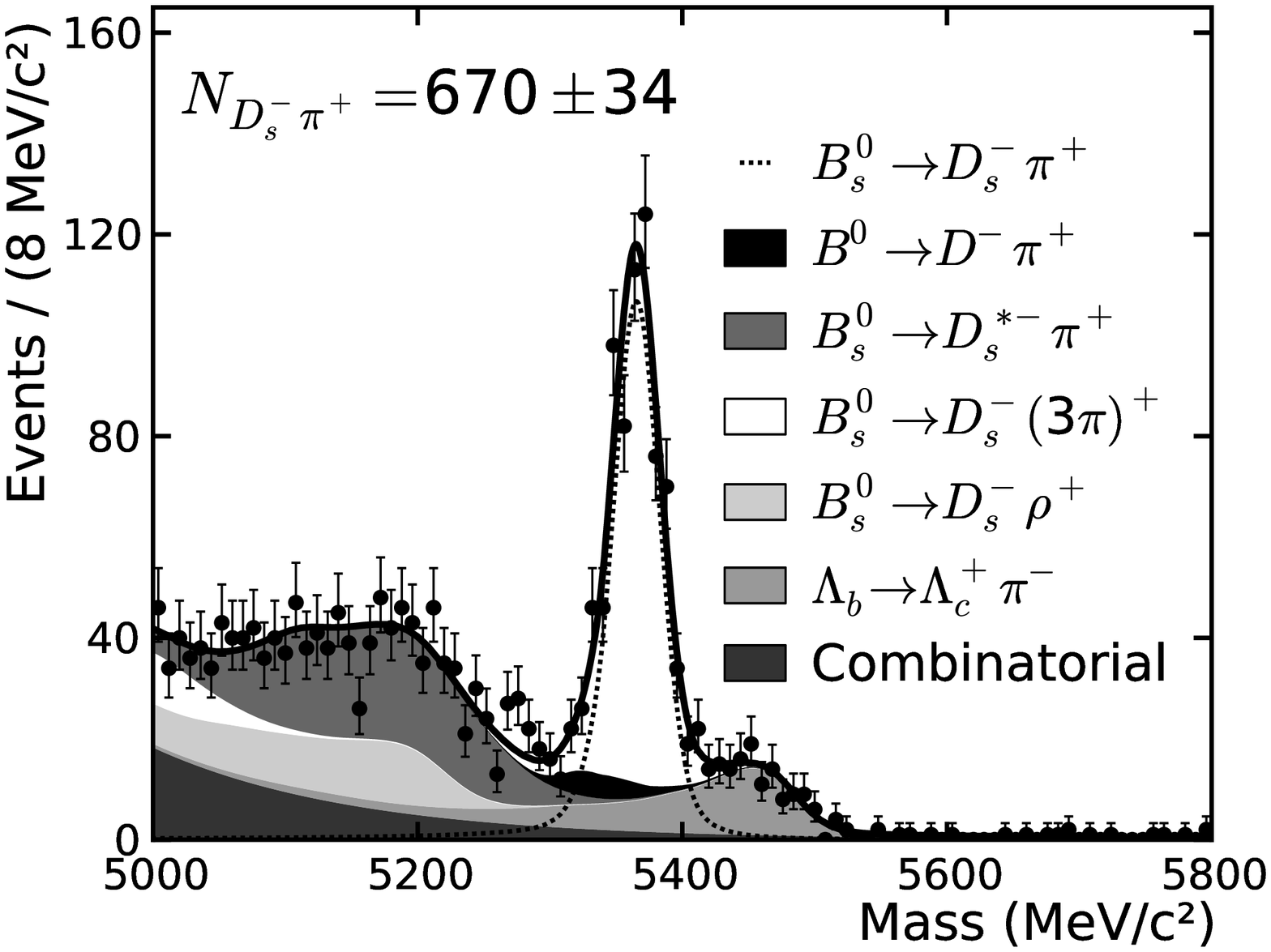}
\caption{
   Mass distributions of the  \BdDp, \BdDK, and \BsDp candidates (top to bottom).
   The indicated components are described in the text.
}
 \label{fig:MisIDCurve_and_DPiFit}
\end{figure}

The relative yields of the three decay modes are extracted from
unbinned extended maximum likelihood fits to the mass distributions
shown in Fig.~\ref{fig:MisIDCurve_and_DPiFit}.
The signal mass shape is described by an empirical model derived
from simulated events.
The mass distribution in the simulation exhibits non-Gaussian tails on either side of the
signal. The tail on the right-hand side is due to non-Gaussian detector effects and modeled
with a Crystal Ball (CB) function
~\cite{Skwarnicki:1986_notitle}. A similar
tail is present on the left-hand side of the peak. In addition, the low mass tail
contains a second contribution
due to events where hadrons have radiated photons that are not reconstructed.
The sum of these contributions is modeled with a second CB function.
The peak values of these two CB functions are constrained to be identical.

Various backgrounds have to be considered, in particular
the crossfeed between the \Dm and \Dsm channels, and the contamination in
both samples from \LbLcp decays, where $\Lambda_c^+ \to pK^-\pi^+$.
The \Dsm contamination in the \Dm data sample is reduced by loose PID requirements,
$\textrm{DLL}(K-\pi)<10$  (with an efficiency of $98.6\%$) and
$\textrm{DLL}(K-\pi)>0$ (with an efficiency of $95.6\%$), for the pions
and kaons from \D decays, respectively. 
The resulting efficiency to reconstruct \BsDp as background
is evaluated, using simulated events, to be 30 times smaller than for \BdDp and
150 times smaller than for \BdDK within the \Bd and \Dm signal mass windows.
Taking into account the lower production fraction of \Bs mesons,
this background is negligible.

The contamination from $\Lambda_c$ decays is estimated in a similar way.
However, different approaches are used for the \Bd and \Bs
decays. A contamination of approximately $2\%$ under the \BdDp mass peak and below
$1\%$ under the \BdDK peak is found, and therefore no explicit
$\textrm{DLL}(p-\pi)$ criterion is needed. The $\Lambda_c$ background in the \Bs
sample is, on the other hand, large enough that it can be fitted for directly.

\begin{figure}[!t]
 \centering
 \includegraphics[width=.40\textwidth]{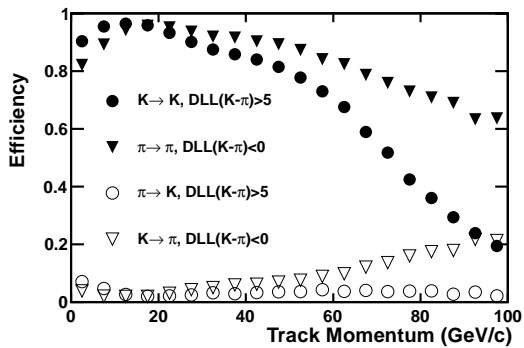}
 \caption{
 Probability, as a function of momentum, to correctly identify (full symbols) a kaon
 or a pion when requiring $\textrm{DLL}(K-\pi)>5$ or $\textrm{DLL}(K-\pi)<0$, respectively.
 The correspondent probability to wrongly identify (open symbols) a pion as a kaon, 
 or a kaon as a pion is also shown.
 The data are taken from a calibration
 sample of $D^*\to D(K\pi)\pi$ decays; the statistical uncertainties are
 too small to display.}
 \label{fig:DPiSigShape_and_MassReweight}
\end{figure}

A prominent peaking background to \BdDK is \BdDp,
with the pion misidentified as a kaon. The small $\pi\to K$
misidentification rate (of about $4\%$) is compensated by the larger branching
fraction, resulting in similar event yields.
This background is modeled by obtaining a clean \BdDp sample from the data
and reconstructing it under the \BdDK mass hypothesis. 
The resulting mass shape depends on the momentum distribution of the bachelor particle.
The momentum distribution after the $\textrm{DLL}(K-\pi)>5$ requirement can be found by
considering the PID performance as a function of momentum.
This is obtained using a sample of $\D^{*+} \rightarrow D^0 \pi^{+} $decays,
and is illustrated in Fig.~\ref{fig:DPiSigShape_and_MassReweight}.
The mass distribution is reweighted using this momentum distribution to reproduce the
\BdDp mass shape following the DLL cut.

The combinatorial background consists of events with random pions and
kaons, forming a fake \Dm or \Dsm candidate, as well as 
real, \Dm or \Dsm mesons that combine with a random pion or kaon.
The combinatorial background is modeled with an exponential shape.

Other background components originate from partially reconstructed
\Bd and \Bs decays.
In \BdDp these originate from
\BdDstarp and \BdDrho decays, 
which can also be backgrounds for \BdDK in the case of a misidentified
bachelor pion. In \BdDK there is additionally background from \BdDstarK decays.
The invariant mass distributions for the partially reconstructed and misidentified
backgrounds are taken from large samples of simulated events, reweighted
according to the mass hypothesis of the signal being fitted and the DLL cuts.

For \BsDp, the \BdDp background peaks under the signal with a similar shape.
In order to suppress this peaking background, PID requirements are
placed on both kaon tracks. The kaon which has the same sign in the \BsDp and \BdDp decays
is required to satisfy $\textrm{DLL}(K-\pi)>0$, while the other kaon in the \Ds decay is
required to satisfy $\textrm{DLL}(K-\pi)>5$.
Because of the similar shape, a Gaussian constraint is applied to the yield of this background.
The central value of this constraint is computed
from the $\pi\to K$ misidentification rate.
The \LbLcp background shape is obtained from simulated events, reweighted according to the 
PID efficiency, and the yield allowed to float in the fit. 
Finally, the relative size of the \BsDrho and \BsDstarp backgrounds is
constrained to the ratio of the \BdDrho and \BdDstarp backgrounds in the \BdDp
fit, with an uncertainty of $20\%$ to account for potential SU(3) symmetry breaking effects.

The free parameters in the likelihood fits to the mass distributions are the event
yields for the different event types, \ie~the combinatorial background,
partially reconstructed background, misidentified contributions, the signal, as well
as the peak value of the signal shape. 
In addition the combinatoric background shape is left free in the \BdDp and \BsDp fits,
and the signal width is left free in the \BdDp fit. In the \BsDp and \BdDK fits the signal
width is fixed to the value from the \BdDp fit, corrected by the ratio of the signal widths
for these modes in simulated events.

The fits to the full \BdDp, \BdDK, and \BsDp data samples are shown in
Fig.~\ref{fig:MisIDCurve_and_DPiFit}.
The resulting \BdDp and \BdDK event yields are $4103\pm 75$ and $252\pm 21$, respectively.
The number of misidentified \BdDp events under the \BdDK signal as obtained from the fit is $131\pm19$.
This agrees with the number expected from the total number of \BdDp
events, corrected for the misidentification rate determined from
the PID calibration sample, of $145\pm5$. The \BsDp event yield is $670\pm 34$.

The stability of the fit results has been investigated using different
cut values for both the PID requirement on the bachelor particle and for
the multivariate selection variable. In all cases variations are found to
be small in comparison to the statistical uncertainty.

The relative branching fractions are obtained by correcting the event yields
by the corresponding efficiency factors;
the dominant correction comes from the PID efficiency.
The dominant source of systematic uncertainty is the knowledge on the \BdDp
branching fraction (for the \BdDK branching fraction measurement) and the knowledge
of the \Dm and \Dsm branching fractions (for the \fsfdt measurement).  
An important source of systematic uncertainty is the knowledge of the PID efficiency as a
function of momentum, which is needed to reweight the mass distribution of the \BdDp
decay under the kaon hypothesis for the bachelor track.  This enters
in two ways: firstly as an uncertainty on the correction
factors, and secondly as part of the 
systematic uncertainty, since the shape for the misidentified backgrounds
relies on correct knowledge of the PID efficiency as a function of momentum.

The performance of the PID calibration is evaluated by applying the same method
from data to simulated events, and the maximum discrepancy found between the 
calibration method and the true mis-identification is attributed as a systematic uncertainty.
The \fsfdt measurement using \BdDK and \BsDp is more robust against PID uncertainties,
since the final states have the same number of kaons and pions.

Other systematic uncertainties are due to limited simulated event samples (affecting
the relative selection efficiencies), neglecting 
the $\Lambda_b \to \Lambda_c^+ \pi^-$ and \BsDp backgrounds in the \BdDp fits,
and the limited accuracy of the trigger simulation. 
Even though the ratio of efficiencies is statistically consistent
with unity, the maximum deviation is conservatively assigned as a
systematic uncertainty. 
The difference in interaction probability between kaons and pions is estimated using MC simulation. 
The systematic uncertainty due to possible discrepancies between data and simulation is expected to 
be negligible and it is not taken into account. 
The efficiency of the non-resonant $D_s$ decays varies across the Dalitz plane, 
but has a negligible effect on the total $B_s^0 \rightarrow D_s^{-} \pi^{+}$ efficiency. 
The sources of systematic uncertainty are summarized in
Tab.~\ref{t:syst_all}.

The efficiency corrected
ratio of \BdDp and \BdDK yields
is combined with the world average of the \BdDp~\cite{PDG_2010_notitle}
branching ratio to give
\begin{equation}
\BR\left(\BdDK\right) = (2.01 \pm 0.18 \pm 0.14)\times 10^{-4}.
\end{equation}
The first uncertainty is statistical and the second systematic.
  \begin{table}
	\begin{center}
   \label{t:syst_all}	
    \caption{Experimental systematic uncertainties for the $\BR\left(\BdDK\right)$ and the two \fsfdt measurements.}
  \begin{tabular}{ lcc }
    \hline
    & $\BR\left(\BdDK\right)$ & \fsfdt \\
    \hline
    PID calibration & $2.5\%$  & $1.0\%$/$2.5\%$ \\
    Fit model & $2.8\%$ & $2.8\%$ \\
    Trigger simulation & $2.0\%$ & $2.0\%$ \\
    $\BR(\BdDp)$  & $4.9\%$ & \\
    $\BR(D^+_s\rightarrow K^+K^-\pi^+)$  & & $4.9\%$ \\
    $\BR(D^+\rightarrow K^-\pi^+\pi^+)$  & & $2.2\%$ \\
    $\tau_{B_s}/\tau_{B_d}$ & & $1.5\%$ \\
    \hline
  \end{tabular}
	\end{center}
  \end{table}

The theoretically cleaner measurement of \fsfdt uses \BdDK and \BsDp
and is made according to Eq.~\ref{eq:fs-det1}.
Accounting for the exclusive \D branching fractions
$\BR(D^+\to K^-\pi^+\pi^+)=(9.14\pm 0.20)\%$~\cite{CleoBRD_notitle} and 
$\BR(D_s^+\to K^-K^+\pi^+)=(5.50\pm 0.27)\%$~\cite{CLEO:2008cqa_notitle},
the value of \fsfdt is found to be
\begin{equation}
\fsfdt = (0.310 \pm 0.030^{\textrm{stat}} \pm 0.021^{\textrm{syst}}) \times  \frac{1}{{\cal N}_a {\cal N}_F},
\end{equation}
where the first uncertainty is statistical and the second is systematic. 
The statistical uncertainty is dominated by the
yield of the \BdDK mode.

The statistically more precise but theoretically less clean
measurement of \fsfdt uses \BdDp and \BsDp and is, from Eq.~\ref{eq:fs-det2},
\begin{equation}
\fsfdt = (0.307 \pm 0.017^{\textrm{stat}} \pm 0.023^{\textrm{syst}} )
\times \frac{1}{{\cal N}_a {\cal N}_F {\cal N}_E}.
\end{equation}
The two values for \fsfdt can be combined into a single value, taking all correlated uncertainties
into account and using the theoretical inputs accounting 
for the $SU(3)$ breaking part of the form factor ratio, the
non-factorizable and W-exchange diagram:
\begin{equation}
\fsfdt = 0.253 \pm 0.017^{\textrm{stat}} \pm 0.017^{\textrm{syst}} \pm 0.020^{\textrm{theor}}.
\end{equation}

In summary, with $35$~pb$^{-1}$ of data collected using the \lhcb detector during the 2010 \lhc operation at
a centre-of-mass energy of 7~TeV, the branching fraction of the Cabibbo-suppressed \Bd decay
mode \BdDK has been measured with better precision than the current world average.
Additionally, two measurements of the \fsfdt production fraction are performed from the relative yields of \BsDp with respect to
\BdDK and \BdDp. 
These values of \fsfdt are numerically close to the values determined at LEP and at the Tevatron~\cite{HFAG_notitle}. 

\section*{Acknowledgments}
\label{sec:acknowledgments}
We express our gratitude to our colleagues in the CERN accelerator departments for the excellent performance of the LHC.
We thank the technical and administrative staff at CERN and at the LHCb institutes, and acknowledge support from the National Agencies: CAPES, CNPq, FAPERJ and FINEP (Brazil); CERN; NSFC (China); CNRS/IN2P3 (France); BMBF, DFG, HGF and MPG (Germany); SFI (Ireland); INFN (Italy); FOM and NWO (Netherlands); SCSR (Poland); ANCS (Romania); MinES of Russia and Rosatom (Russia); MICINN, XUNGAL and GENCAT (Spain); SNSF and SER (Switzerland); NAS Ukraine (Ukraine); STFC (United Kingdom); NSF (USA).
We also acknowledge the support received from the ERC under FP7 and the R\'{e}gion Auvergne. 
\bibliographystyle{unsrt}
\bibliography{fdfsreferences}

\clearpage
\center{\bf{The LHCb Collaboration}}
\onecolumngrid 
\begin{flushleft}
\small
\begin{centering}
R.~Aaij$^{23}$, 
B.~Adeva$^{36}$, 
M.~Adinolfi$^{42}$, 
C.~Adrover$^{6}$, 
A.~Affolder$^{48}$, 
Z.~Ajaltouni$^{5}$, 
J.~Albrecht$^{37}$, 
F.~Alessio$^{6,37}$, 
M.~Alexander$^{47}$, 
G.~Alkhazov$^{29}$
% ...
\\

{\it
% ...
$ ^{5}$Clermont Universit\'{e}, Universit\'{e} Blaise Pascal, CNRS/IN2P3, LPC, Clermont-Ferrand, France\\
$ ^{6}$CPPM, Aix-Marseille Universit\'{e}, CNRS/IN2P3, Marseille, France\\
% ...
$ ^{23}$Nikhef National Institute for Subatomic Physics, Amsterdam, Netherlands\\
% ...
$ ^{29}$Petersburg Nuclear Physics Institute (PNPI), Gatchina, Russia\\
% ...
$ ^{36}$Universidad de Santiago de Compostela, Santiago de Compostela, Spain\\
$ ^{37}$European Organization for Nuclear Research (CERN), Geneva, Switzerland\\
% ...
$ ^{42}$H.H. Wills Physics Laboratory, University of Bristol, Bristol, United Kingdom\\
% ...
$ ^{47}$School of Physics and Astronomy, University of Glasgow, Glasgow, United Kingdom\\
$ ^{48}$Oliver Lodge Laboratory, University of Liverpool, Liverpool, United Kingdom\\
% ...
\bf{(The LHCb Collaboration)}\\
}
\end{centering}
\end{flushleft}

\end{document}